\documentclass{article}

\usepackage[final]{neurips_2022}

\usepackage[utf8]{inputenc} 
\usepackage[T1]{fontenc}    
\usepackage{hyperref}       
\usepackage{url}            
\usepackage{booktabs}       
\usepackage{nicefrac}       
\usepackage{microtype}      
\usepackage[dvipsnames]{xcolor}         
\hypersetup{hidelinks}
\usepackage{tikz}
\usepackage{amsmath}
\usepackage{amssymb}
\usepackage{amsthm}
\usepackage{cleveref}
\usepackage{bm}
\usepackage{bbm}
\usepackage{xfrac}
\usepackage{enumitem}
\usepackage{blkarray}
\usepackage{mathtools}
\usepackage{xcolor}
\usepackage{colortbl}
\definecolor{Gainsboro}{RGB}{220,220,220}
\usepackage{array}
\newcolumntype{P}[1]{>{\raggedright\arraybackslash}p{#1}}
\usepackage{makecell}
\newcommand{\redcircle}[2][red,fill=red]{\tikz[baseline=-0.5ex]\draw[#1,radius=#2] (0,0) circle ;}
\newcommand{\blackcircle}[2][black,fill=black]{\tikz[baseline=-0.5ex]\draw[#1,radius=#2] (0,0) circle ;}

\newcommand{\greencircle}[2][green,fill=green]{\tikz[baseline=-0.5ex]\draw[#1,radius=#2] (0,0) circle ;}%

\title{The fitness landscape of social norms\\ in social dilemmas}
\author{%
  Maximilian Puelma Touzel\thanks{\href{http://mptouzel.github.io}{mptouzel.github.io}} \\
  Mila - Quebec AI Institute\\
  Montréal, Canada \\
  \texttt{puelmatm@mila.quebec} \\
}

\begin{document}
\maketitle
\begin{abstract}
By specifying behaviour across multiple agents, social norms are a coordination approach to resolving social dilemmas.  
Decentralized and wide adoption can be achieved by norms whose prescription involves interpreting stochastic signals in the environment. Such signals must have enough \textit{correlation} to orchestrate mutually beneficial coordination and enough disincentivizing \textit{uncertainty} about the benefits of exploiting that coordination. 
Evolutionary game theory of matrix games has been used to describe how, by rational agents comparing and adopting norms, a norm can evolve to become dominant in a population. \cite{Morsky2019} classify norms according to a set of rationality criteria. Joint player strategies that adopt norms that are consistent with optimal single-player strategies with respect to expected reward naturally satisfy a correlated, rather than Nash game theoretic equilibrium condition. Here, we present a version of this theory that clarifies the basic ingredients. We formulate it in the more general Markov game setting more commonly used in reinforcement learning theory. We illustrate the theory by mapping norms over the signal and reward space, while also giving a detailed exposition of the underlying mechanics of the approach. 
Finally, we give a general solution and analysis of replicator dynamics, which \cite{Morsky2019} propose as a means by which these norms could emerge.

\end{abstract}

\vspace{-15pt}
\subsection*{Introduction}

In the epistemic perspective on game theory~\citep{Aumann1987}, uncertainty about opponents’ strategies is framed as an inference problem that must tackled in pursuit of an optimal strategy%
\footnote{Note that optimal strategies for perfect knowledge settings (\textit{e.g.} of other agents) are always pure strategies.}.
Through any consistent calibration process \citep{Foster1997,Hart2000}, each player learns to infer a posterior over the plays of the opponent given the player's observation of the state. The natural Bayes optimal strategy is then for a player to take the action that gives the highest posterior-averaged utility. Surprisingly, this fundamental process does not lead to the celebrated solution concept of a Nash equilibrium\footnote{Which assumes, unrealistically, that the player knows the policies of the opponent.}, but rather to the lesser known concept of a \textit{correlated equilibrium}. 

\cite{Morsky2019} present a theory for how correlated equilibria emerge endogenously through an evolution of social norms. Here, we present a minimal version of their theory (\textit{e.g.} without partitions)  using the notation of reinforcement learning theory and illustrate it with an application to the game of chicken. To demonstrate the theory, \cite{Morsky2019} take the well-established approach to gaining theoretical traction in evolutionary dynamics of game-playing agents by 
considering a dynamics wherein pairs of agents are selected at random for interaction%
\footnote{This strong assumption can be relaxed by moving to a finite population description in which interactions are specified (e.g. assortative or via a given social graph) at the expense of having to resort to numerical analysis of solutions \citep{Hilbe2011}.}.
Under this assumption, a norm selection dynamics can be faithfully represented through a particularly simple mean-field description called replicator dynamics. After presenting our exposition of the theory, we give a complete characterization of the game family of the game of chicken and then go on to give a general formulation and analysis of replicator dynamics.

\subsection*{Related Work}
This work translates ideas at the intersection of evolutionary game theory and sociology to the multi-agent reinforcement learning research field. The concept of correlated equilibria was introduced by the works of \cite{Aumann1987}. Sociologists then proposed correlated equilibrium as a concept central to social norms \citep{Gintis} and game theoretic work on this topic, albeit sparse, extends back decades\citep{Axelrod} (see \cite{Morsky2019}'s account of this literature). In parallel, correlated equilibria have been studied for decades in the multi-agent reinforcement learning (MARL) theory community, with seminal contributions from the group of Amy Greenwald, \textit{e.g.} \cite{Greenwald}.  The recent work in that field is expanding its realm of applicability, e.g. to extensive form games \cite{Anagnostides} (and references therein). Separately and much more recently, the emergence of social norms has been studied empirically using deep reinforcement learning agents, e.g. \cite{Koster} and \cite{Vinitsky}. Our work aims to motivate the intermingling of these empirical and theoretical traditions by developing the evolutionary game theory of social norms as studied in evolution and social science in the setting most relevant to models of multi-deep reinforcement learning agent systems that are currently being used to explore the social dimensions of artificial intelligence.

\subsection*{Background}
\paragraph{Markov Game Social Dilemmas}
We focus on the two-player partially observable Markov game setting. In this simple illustration, we assume state transitions do not depend on the actions and rewards do not depend on the state. We also restrict ourselves to the symmetric case. Accordingly, we denote the player with unprimed variables, while those of the opponent are primed. All sets are discrete.

The environment is a set of states $\mathcal{S}$. State information available to agents is bounded by (possibly stochastic) observation functions $p(o|s)$ and $p(o'|s)$ for observations $o$ and $o'$ from observations spaces $\mathcal{O}$ and $\mathcal{O}'$ for player and opponent, respectively.
Action-independent state transitions are given by $\mathcal{T}(\tilde{s}|s)$ where $\tilde{s}$ is the next state. 
We assume a stationary environment so the state description reduces to the stationary state distribution $p(s)$ defined implicitly by $p(\tilde{s})=\sum_{s} \mathcal{T}(\tilde{s}|s)p(s)$. The (in general state-correlated) joint observation distribution is then $p(o,o')=\sum_s p(o|s)p(o'|s)p(s)$.
For $\bm{p}_s$ denoting the vector representation of $p(s)$ and $\bm{P}_{oo'}$ denoting the matrix representation of $p(o,o')$, we have the matrix expression $\bm{P}_{oo'}=\bm{P}_{o|s}\mathrm{diag}(\bm{p}_s)\bm{P}_{o'|s}^\top$, where  $\bm{P}_{o|s}$ is the $|\mathcal{O}|\times|\mathcal{S}|$ matrix representation of $p(o|s)$ and similarly for $p(o'|s)$\footnote{In this two-player, stochastic policy setting, the matrix algebra of the probability theory of two-dimensional, discrete random variables is useful. 
Namely, for a matrix representation of a joint distribution on $x$ and $y$, $\bm{P}_{xy}$, the marginals are computed as $\bm{p}_x=\bm{P}_{xy}\bm{1}$ and $\bm{p}_y=\bm{P}^\top_{xy}\bm{1}$. Independence is the factorization condition $\bm{P}_{xy}=\bm{p}_x\bm{p}^\top_y$ (when $x$ and $y$ are binary, $\bm{P}_{xy}$ is determined by the marginals and the correlation, $\rho$, $\bm{P}_{xy}=\bm{p}_x\bm{p}_y^\top+\rho\sqrt{\sigma_x^2\sigma_y^2}\begin{bsmallmatrix} +1 &-1 \\-1& +1\end{bsmallmatrix}$, where $\sigma^2_i=\mathrm{det}(\mathrm{diag}(\bm{p}_i))$ for $i=x,y$ is the variance of $x$ and $y$ respectively). The expectation of a scalar function $f(x,y)$ is computed as $\mathbb{E}_{(x,y)}[f(x,y)]=\mathrm{tr}(\bm{F}\bm{P}^\top_{xy})$, where variable $\bm{F}$ is the matrix of $f(x,y)$ function evaluations. 
Adding conditional dependence via a correlated pair of random variables  $(\tilde{x},\tilde{y})$ with joint distribution $\bm{P}_{\tilde{x}\tilde{y}}$ gives $\bm{P}_{xy}=\bm{P}_{y|\tilde{y}}\bm{P}_{\tilde{x}\tilde{y}}^\top\bm{P}_{x|\tilde{x}}^\top$, where $\bm{P}_{y|\tilde{y}}$ and $\bm{P}_{x|\tilde{x}}$ are the matrix representations of the respective conditional distributions.}.
With the state variable marginalized, we hereon represent the environment by $\bm{P}_{oo'}$. 



A player policy (a possibly mixed strategy) $\pi:\mathcal{O}\to \Delta^{|\mathcal{A}|-1}$ is a mapping from the observation space to distributions on an action space, $\mathcal{A}$. $\pi$ can be represented as a  $|\mathcal{A}|\times|\mathcal{O}|$ matrix $\bm{P}_{a|o}$, such that $\sum_i (\bm{P}_{a|o})_{ij}=1$ for all $j$. A deterministic policy (pure strategy) corresponds to policies for which $(\bm{P}_{a|o})_{ij}\in\{0,1\}$. 
The opponent's strategy $\pi'$ is defined similarly.

In practise, $\mathcal{O}$ will be a large space of which only a small part of which will be useful as signal by the agent for any given task. 
Here, we will assume that the agent has performed this selection and that $\mathcal{O}$ is the signal space it attends to get suggestions for how to act. Thus, we assume $|\mathcal{O}|=|\mathcal{A}|$.
To map onto the simply symmetric matrix games discussed below, we identify the two action spaces $\mathcal{A}=\mathcal{A}'$ (and thus also the observation spaces $\mathcal{O}=\mathcal{O}'$). We also assume the observation spaces confer no player bias so the marginals $p(o)$ and $p(o')$ are equal (achieved by the symmetry constraint, $\bm{P}_{oo'}=\bm{P}_{oo'}^\top$). 

A reward function in this symmetric player setting is a mapping $r:\mathcal{A}\times\mathcal{A}'\to \mathbb{R}$ returning the reward to a player 
for playing action $a_i\in\mathcal{A}$ when its opponent plays $a'_j\in\mathcal{A}'$. We can represent $r$ using the $|\mathcal{A}|\times|\mathcal{A}|$ action reward matrix $\bm{R}$ where $r(a=a_i,a'=a_j)=R_{ij}$. Note that unlike a general Markov game, the reward here does not depend on the environment observations. 

Canonical social dilemmas are two-player matrix games with a binary action space corresponding to a cooperative and exploitative action. The reward matrix for such dilemmas is
\begin{align}
\bm{R}=
\begin{bmatrix}
    B & S\\
    T & P
\end{bmatrix}
 \;\textrm{with}&\;\;
\begin{tabular}{@{}l@{}}
$B$, the benefit for mutual cooperation; \\
$P$, the punishment for mutual exploitation;\\
$S$, the sucker's payoff for cooperating when the other has exploited; \&\\ 
$T$, the tempting reward for exploiting when the other has cooperated. \end{tabular}
\end{align}
In this setting, the conditions for a tension between individual and collective interests that characterize a social dilemma \citep{Macy2002} are that
\begin{enumerate}[nosep]
    \item \;\;\;$B>P$: mutual cooperation > mutual exploitation
    \item \;\;\;$B>S$: mutual cooperation > being suckered
    \item \;\;$2B>T+S$: mutual cooperation > equal mix of unilateral cooperation and exploitation  
    \item a)\;\;$T>B$ (\textit{greed}): exploiting a cooperator > mutual cooperation\par \textit{and/or} \par
        b)\; $P>S$ (\textit{fear}): mutual exploitation > being exploited. 
\end{enumerate}
Condition 3 ensures cooperation offers more than pure competition (\textit{i.e.} to exploit and be exploited). Condition 4 gives 3 classes of dilemmas
\begin{center}
    \begin{tabular}{r|c|c}
        & not \textit{greed} & \textit{greed}\\
        \hline
        not \textit{fear} & \begin{tabular}{@{}c@{}}no dilemma\\$B>T>S>P$\end{tabular} &\begin{tabular}{@{}c@{}}\textit{game of chicken} \\ $T>B>S>P$\end{tabular}\\
        \hline
        \textit{fear} & \begin{tabular}{@{}c@{}}\textit{stag hunt}\\ $B>T>P>S$\end{tabular} & \begin{tabular}{@{}c@{}}\textit{prisoner's dilemma}\\ $T>B>P>S$\end{tabular}
    \end{tabular}
\end{center}
These conditions have been applied to the sequential decision-making setting relevant to reinforcement learning to try to leverage the insight they have offered into matrix games \citep{Leibo2017}. We will focus on the game of chicken since it serves as the canonical example of correlated equilibria.

\paragraph{Game of chicken example} In the case of \textit{greed} and not \textit{fear}, $T>B>S>P$ characterizes the canonical family of anti-coordination social dilemmas known as the \textit{game of chicken} where in one interpretation, $\mathcal{A}=\{stop,go\}$ defines the actions available to each of two cars approaching an intersection from different roads. This game family captures the inevitability of conflict in the pursuit of self-interest as a coordination dilemma: highest reward for \textit{go}, but only when the other chooses to \textit{stop}, otherwise the reward is low for both agents (because they crash into eachother). The car interpretation evokes the coordination solutions (traffic lights, the \textit{yield-to-the-right} rule, etc.) that have evolved to facilitate the resolution of this dilemma. For this reason, the game is often invoked as a simple setting to explain a correlated equilibrium, a solution concept by which social dilemma are rationally resolved. In this case, the game environment is augmented with observation spaces ($\mathcal{O}=\mathcal{O}'=\{\redcircle{2pt},\greencircle{2pt}\}$ in the traffic light solution) that provides signals to be interpreted as suggested actions (e.g. the signal-following norm: to stop when \redcircle{2pt} is observed, and go when \greencircle{2pt} is observed). There are multiple ways to represent this simple matrix game in the more general Markov game setting outlined above. Here, we represent the game by defining observation functions as partitions on the state space such that $(\bm{P}_{o|s})_{ij},(\bm{P}_{o'|s})_{ij}\in\{0,1\}$. The state space then naturally partitions into the refined partition obtained by intersecting the two player's partitions. This refined partition abstracts $\mathcal{S}$ into $|\mathcal{A}|^2=4$ states, $\{(\redcircle{2pt},\redcircle{2pt}),(\redcircle{2pt},\greencircle{2pt}),(\greencircle{2pt},\redcircle{2pt}),(\greencircle{2pt},\greencircle{2pt})\}$. In a minimal example, $|\mathcal{S}|=4$ and partitions can be chosen such that the elements of $\bm{P}_{oo'}$ are those of $\bm{p}_s=(p_1,p_2,p_3,p_4)$. For example,
\begin{equation}
\bm{P}_{o|s}=\begin{bmatrix}
    1& 1& 0& 0 \\0& 0& 1& 1
\end{bmatrix}\;\;,\;\;
\bm{P}_{o'|s}=\begin{bmatrix}
    1 &0 &1& 0 \\0& 1& 0& 1
\end{bmatrix}\;\;\implies
\bm{P}_{oo'}=\begin{bmatrix}
   p_{{\redcircle{1pt}}\redcircle{1pt}}&p_{{\redcircle{1pt}}\greencircle{1pt}}\\
   p_{{\greencircle{1pt}}\redcircle{1pt}}&p_{{\greencircle{1pt}}\greencircle{1pt}}
\end{bmatrix}=\begin{bmatrix}
   p_1&p_2\\
   p_3&p_4
\end{bmatrix}\;.
\end{equation}
The symmetry assumption on $\bm{P}_{oo'}$, $ p_{{\greencircle{1pt}}\redcircle{1pt}}= p_{{\redcircle{1pt}}\greencircle{1pt}}$, demands that $p_2=p_3$. Having control over $\bm{P}_{oo'}$ here means having control over $\bm{p}_s$. However, for the general case of a large $\mathcal{S}$, given a fixed distribution $\bm{p}_s$, control over $\bm{P}_{oo'}$ could also be achieved via the freedom in partitioning through the joint sculpting of $\bm{P}_{o|s}$ and $\bm{P}_{o'|s}$.
%

\paragraph{Correlated equilibrium} Let us denote by $S$ the system without the player's policy, $S=(\pi',p(o,o'),r)=(\bm{P}_{a'|o'},\bm{P}_{oo'},\bm{R})$ . The \textit{average reward} $\rho$ for the player is then%
\begin{align}
    \rho(\pi,S)=\mathbb{E}[r(a,a')|\pi,S]:&=\sum_{o,o',a,a'}r(a,a')\pi'(a'|o')p(o,o')\pi(a|o)\\
    &=\mathrm{tr}\left(\bm{R}\bm{P}_{a'|o'}\bm{P}_{oo'} \bm{P}_{a|o}^\top\right),\label{eq:matrho}
\end{align}
where we omit the transpose on $\bm{P}_{oo'}$ since it is symmetric. $\rho(\pi',S')$ is defined similarly where $p'$ is $p$ with swapped arguments ($p'=p$ for the unbiased observation distributions we consider here) and $r'=r$ for this symmetric player setting.
A policy $\bm{P}_{a|o}$ that maximizes $\rho$, \textit{i.e.} $\mathrm{tr}(\bm{R}\bm{P}_{a'|o'}\bm{P}_{oo'} (\bm{P}_{a|o}-\tilde{\bm{P}}_{a|o})^\top)\geq 0$ for any other policy $\tilde{\bm{P}}_{a|o}$, is best in expectation over all randomness including over the observation, $o$, observed by the player. A stronger notion of optimality is that $\bm{P}$ is optimal for \textit{each} observation value $o=o_i$, $((\bm{R}\bm{P}_{a'|o'}\bm{P}_{oo'})_{\cdot i})^\top(\bm{P}_{a|o}-\tilde{\bm{P}}_{a|o})_{\cdot i}\geq 0$ for all $i$\footnote{$\mathrm{diag}((\bm{R}\bm{P}_{a'|o'}\bm{P}_{oo'})(\bm{P}_{a|o}-\tilde{\bm{P}}_{a|o})^\top)\geq \bm{0}$ gives the matrix form of the set of $|\mathcal{O}|$ inequalities (here the function $diag$ obtains the diagonal of its matrix argument).}. Such a policy is called a \textit{best response} (BR) policy. 

More generally, we can think of a (not necessarily factorizable) joint action distribution $p(a,a')=\bm{P}_{a|o}\bm{P}_{oo'} \bm{P}_{a'|o'}^\top$ and the environmental observations simply as a means to generate action-action correlations between the player and its opponent. In this context, a \textit{correlated equilibrium} given $r$ is the observation ensemble $p$ and a pair of deterministic policies $(\pi,\pi')$ that are best responses to the expected behaviour of the other given their observations, \textit{i.e.} 
for $\tilde{\pi}$ and $\tilde{\pi}'$ denoting any other strategy for player and opponent, respectively, 
\begin{equation}
    \begin{array}{rll}
    ((\bm{R}\bm{P}_{a'|o'}\bm{P}_{oo'})_{\cdot i})^\top(\bm{P}_{a|o}-\tilde{\bm{P}}_{a|o})_{\cdot i}\geq 0
    & \textrm{from the player's perspective }S\; \textrm{, and}\\ 
    ((\bm{R}\bm{P}_{a|o}\bm{P}_{oo'})_{\cdot i})^\top(\bm{P}_{a'|o'}-\tilde{\bm{P}}_{a'|o'})_{\cdot i}\geq 0
    & \textrm{from the opponent's perspective }S'\;.\label{eq:correqdef}
\end{array}
\end{equation}
A \textit{Nash equilibrium} is the special case where the joint action distribution factorizes, $p(a,a')=p(a)p(a')$, \textit{i.e.} the factorization condition $\bm{P}_{a|o}\bm{P}_{oo'} \bm{P}_{a'|o'}^\top=\bm{p}_a\bm{p}_{a'}^\top$ into the outer product of its marginal distribution vectors. 

\subsection*{Social norm theory}
A social norm is a belief to which an agent can subscribe that characterizes a social interaction involving multiple agents. For a pair-agent interaction, a social norm for a player both \textit{prescribes} to that player how they should act in that interaction (via the prescribed policy, $\pi_{\mathrm{prescribed}}$, here represented by $\bm{P}_{a|o}$), and \textit{describes} to that player how the opponent should act in the same interaction (the described policy $\pi_{\mathrm{described}}$, here represented by a policy matrix denoted $\bm{D}_{a'|o'}$). 
A norm as a belief within a formal reasoning framework is thus represented as the matrix pair $(\bm{P}_{a|o},\bm{D}_{a'|o'})$. These are deterministic assignments so $\bm{P}_{a|o}$ and $\bm{D}_{a'|o'}$ are binary-valued matrices with a single non-zero entry per column. The total number of norms is then equal to the total number of unique ordered pairs of policies, 
$N=(|\mathcal{A}||\mathcal{O}|)^2$ norms.

\begin{figure}[tpb!]
\vskip 0.2in
\begin{center}
\centerline{\includegraphics[width=0.5\textwidth]{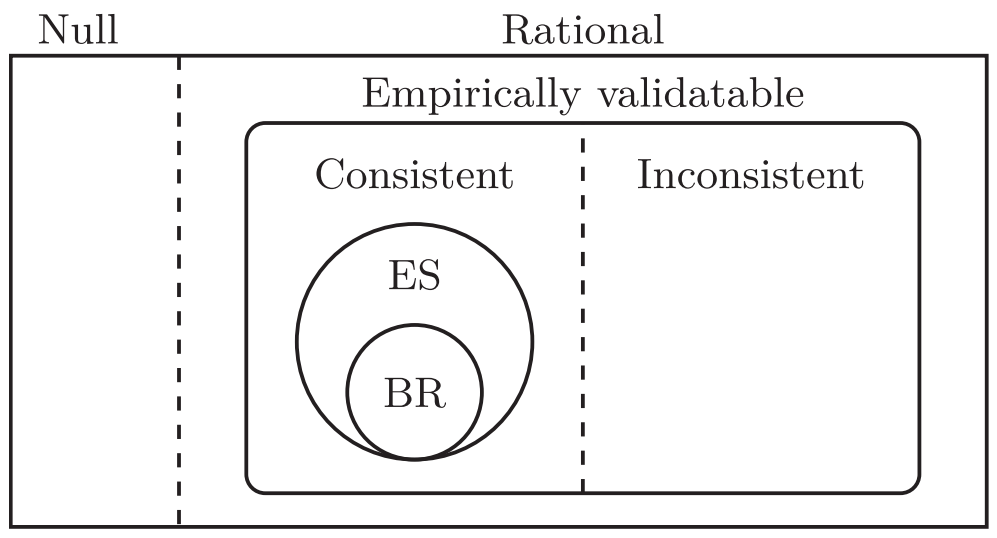}
}
\caption{\textbf{Classes of norms}. Diagram taken from \cite{Morsky2019}. 
}
\label{fig:norm_classes}
\end{center}
\vskip -0.2in
\end{figure}

Let $n,n'\in\{1,\dots, N\}$ index these pairs so that the average reward $\Gamma_{nn'}$ (hereon called the norm payoff) for a player playing within a norm $n$ that indexes $(\bm{P}_{a|o},\bm{D}_{a'|o'})$ against an opponent playing within the norm $n'$ that indexes $(\bm{P}'_{a|o},\bm{D}'_{a'|o'})$ is just $\rho(\pi,S)$ (\cref{eq:matrho}) with $\bm{P}_{a'|o'}=\bm{P}'_{a|o}$. 
The norm payoff matrix is then denoted $\bm{\Gamma}=(\Gamma_{nn'})$. While the descriptive part of a norm $\bm{D}_{a'|o'}$ does not affect its payoff directly, it does affect whether it is rational for a player to adopt the norm.  \cite{Morsky2019} provide a classification of norms via increasingly strong notions of validity (\textit{c.f.} \cref{fig:norm_classes}). Namely, \textit{rational} norms are those that are internally valid: the prescription $\bm{P}_{a|o}$ is never worse than any other prescription $\tilde{\bm{P}}_{a|o}$ given the description, $\bm{D}_{a'|o'}$, else they are called \textit{null} norms. Null norms are never optimal so in that case the player plays a (observation-independent) default strategy instead (\cite{Morsky2019}'s choice being that the player eschews null norms completely and plays the best Nash strategy). This default strategy competes with the prescriptions of norms in any selection dynamics. Norms that are externally valid are called \textit{empirically validatable}: the prescription is rational against the prescription of another rational norm. If an empirically validatable norm is validated by itself, it is called \textit{consistent}, else it is \textit{inconsistent}. Within the set of consistent norms is the set of norms that are \textit{evolutionarily stable} (ES), within which are the set of \textit{best response} (BR) norms. 

From this norm classification, we can situate correlated equilbria: a pair of norms, $n$ and $n'$, form the correlated equilibrium $\bm{P}_{oo'}$ and $(\bm{P}_{a|o},\bm{P}'_{a|o})$ if $n$ and $n'$ are empirically validatable with respect to one another, \textit{i.e.}
 \cref{eq:correqdef}, given the matrix representation of norms.
Under a natural selection process, the observed set of correlated equilibria are those arising from evolutionarily stable norms, \textit{i.e.} norms that are stable to perturbations in norm space. \cite{Morsky2019} prove that such norms must be consistent, \textit{i.e.} their prescriptions are best-responses to the opponent-environment system formed by thier descriptions, and vice versa. Next, we give an example application of this social norm theory for the game of chicken and then go on to describe an natural selection dynamics among game-playing agents by which a norm implementing a correlated equilibrium emerges.

\subsection*{Theory application to the canonical coordination game family: the game of chicken}
The game of chicken constraints on $\bm{R}$ and the linearity of our objective (expectations of $\bm{R}$), as well as the normalization and symmetry constraints on $\bm{P}_{oo'}$ allow for a complete characterization of this canonical system.

By the linearity of the operations (taking expectations) that we will perform on $\bm{R}$, $P$ and $S$ can be set without loss of generality as the reference and unit of reward, respectively, so that $P=0$ and $S=1$. Condition 2 ($B>S$) is then $B>1$. From the greed condition 4a ($T>B$), we can parametrize temptation by the excess reward $L=T-B>0$ as the strength of \textit{greed} (\textit{c.f.} \cite{Leibo2017}). The reward matrix is then expressed
\begin{equation}
\bm{R}(B,L)=
\begin{blockarray}{rcc}
a\backslash a'&stop & go \\
\begin{block}{r[cc]}
  stop & B & 1 \\
  go & B+L & 0 \\
\end{block}
\end{blockarray}\label{eq:rewardmtr}
\end{equation}
with $B>1$, $L>0$, and condition 3 implying, $L<B-1$, so that the exploitation excess can not overwhelm mutual cooperation\footnote{The prisoner's dilemma family can be obtained from this matrix by swapping the elements in the second column and for the condition $L<B$.}. We call $L$ the \textit{coordination challenge level} since it sets the strength of temptation relative to that of coordination.

\begin{figure}[tpb!]
\vskip 0.2in
\begin{center}
\centerline{\includegraphics[width=\textwidth]{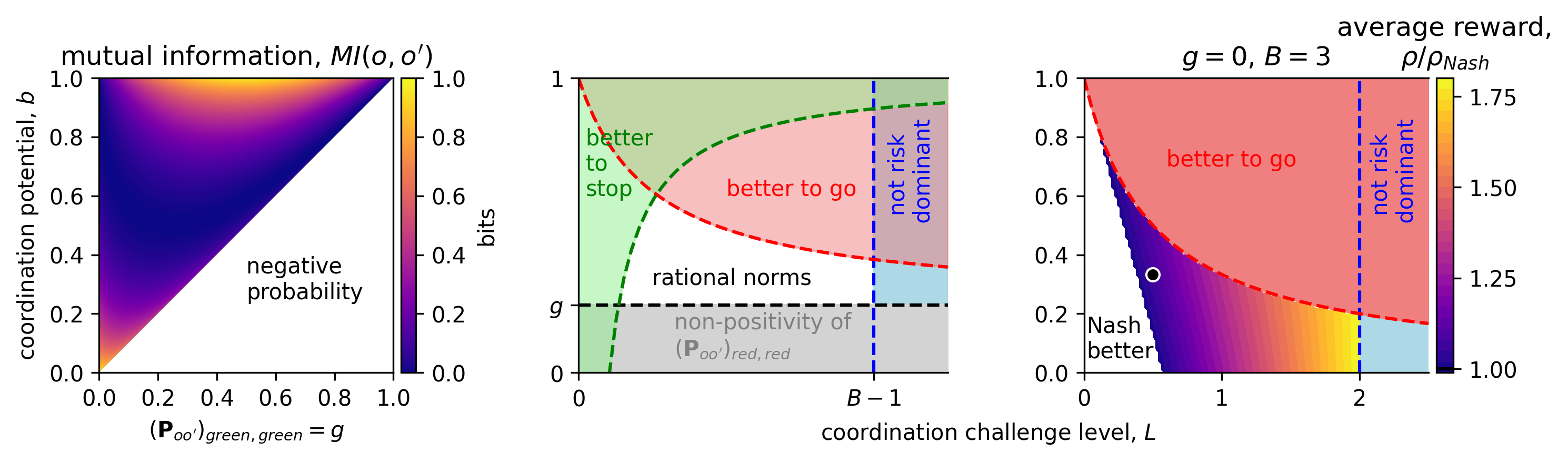}
}
\caption{\textbf{Phase diagram for the rationality of the signal-following norm over the game family of the game of chicken}. \textit{Left}: mutual information between environmental signals $o$ and $o'$ as a function the parametrization of $\bm{P}_{oo'}$, $(b,g)$. \textit{Center}: phase diagram for the general case. The norm is not always rational: when \redcircle{2pt} is observed, it is better to $go$ when $b>(1+2gL)/(1+2L)$ (red). Similarly, when \greencircle{2pt} is observed, it better to stop when $b>1-2g/L$ (green). Positivity of $p_{\redcircle{1pt}\redcircle{1pt}}$ requires $b>g$ (gray). When $L>B-1$ (blue), pure competition (to exploit and be exploited) is more advantageous (\textit{c.f.} social dilemma condition 3). \textit{Right}: Case of $g=0$ and $B=3$ with average reward relative to that of the mixed Nash equilibrium (colorbar). The x-axis, $L$, sets the hardness of the coordination problem and the y-axis, $b$, sets the coordination potential. Harder coordination problems limit the amount of rational cooperation. 
}
\label{fig:phasediagram}
\end{center}
\vskip -0.2in
\end{figure}

Under the signal-following norm, $\bm{P}_{oo'}$ serves as the joint action distribution. As a result, we can parametrize it in a way that highlights performance. Its symmetry and the normalization constraint leave it with two degrees of freedom. Given the payoff matrix \cref{eq:rewardmtr}, the action pair, $(stop,stop)$, gives the highest average reward (by construction, \textit{c.f.} social dilemma condition 3). This is the case where the two players fully coordinate. We thus define the \textit{coordination potential}, $b$ and set $p_{{\redcircle{1pt}}\redcircle{1pt}}\propto b>0$. Realizing this potential is limited by exploitation, which we parametrize using $g=p_{{\greencircle{1pt}}\greencircle{1pt}}$, and set $p_{{\redcircle{1pt}}\redcircle{1pt}}\propto -g$. 
This gives the parametrization
\begin{equation}
\bm{P}_{oo'}(b,g)=
\begin{blockarray}{ccc}
o\backslash o' &\redcircle{2pt} & \greencircle{2pt} \\
\begin{block}{c[cc]}
 \redcircle{2pt}  & b-g&(1-b)/2\\
\greencircle{2pt}& (1-b)/2&g\\
\end{block}
\end{blockarray}\label{eq:Ab}\;
\end{equation}
with $b>g$ by the positivity of  $p_{{\redcircle{1pt}}\redcircle{1pt}}$.
The goal then is to maximize $b-g$ under the constraint that it does not incentivize exploitation. There is no protection from exploitation from placing mass on the go-go signal so $g$ can be optimized independently by making it as small as possible ($g=0$), leaving the problem to maximize $b$ subject to the constraint. Since $b<1$ leaks probability into the dislike pairs, coordination can be made rational even when eliminating the undesirable $(go,go)$ action pair with $g=0$ (we analyze this desirable case below). The pairwise correlation is $2b-1$ and the mutual information (see \cref{fig:phasediagram}a) achieves the maximum of 1 bit at the pure symmetric $(b,g)=(1,1/2)$ and pure antisymmetric $(b,g)=(0,0)$ configurations. 

Using the above parametrizations of $\bm{R}$ and $\bm{P}_{oo'}$, the task-environment can be completely characterized within the plane of coordination potential $b$ and coordination challenge level $L$. Here, we give an example for the signal-following norm (\textit{i.e} \textit{stop} when \redcircle{2pt}; \textit{go} when \greencircle{2pt}). We first derive the norm's rationality conditions. There is a best-response (BR) condition for each of the signal values:
\begin{center}
\begin{tabular}{c|c|c|c}
\begin{tabular}{@{}c@{}}received\\ signal probability\end{tabular} &
\begin{tabular}{@{}c@{}}posterior of \\opp.'s signal\\$p_{i|j}=p(o'=i|o=j)$\end{tabular} & 
\begin{tabular}{@{}c@{}}reward for\\playing action $a$ when opp.\\follows norm ($a'\sim o'$) \end{tabular} &
\begin{tabular}{@{}c@{}}rational: \\$o$ is BR \end{tabular}\\
\hline
\begin{tabular}{@{}c@{}}$p(o=\greencircle{2pt})$ \\$=g+(1-b)/2$\end{tabular} &
$p_{{\redcircle{1pt}}|\greencircle{1pt}}=\frac{(1-b)/2}{g+(1-b)/2}$ &  
\begin{tabular}{@{}l@{}}$stop$: \;\;\;\;\;\;$Bp_{{\redcircle{1pt}}|\greencircle{1pt}}+1(1-p_{{\redcircle{1pt}}|\greencircle{1pt}})$ \\$go$: $(B+L)p_{{\redcircle{1pt}}|\greencircle{1pt}}+0(1-p_{{\redcircle{1pt}}|\greencircle{1pt}})$\end{tabular} & 
$b<1-\frac{2g}{L}$\\
\hline
\begin{tabular}{@{}c@{}}$p(o=\redcircle{2pt})$ \\$=b-g+(1-b)/2$\end{tabular} & 
$p_{{\redcircle{1pt}}|\redcircle{1pt}}=\frac{b-g}{b-g+(1-b)/2}$ & 
\begin{tabular}{@{}l@{}}
    $stop$: \;\;\;\;\;\;$Bp_{{\redcircle{1pt}}|\redcircle{1pt}}+1(1-p_{{\redcircle{1pt}}|\redcircle{1pt}})$\\
    $go$: $(B+L)p_{{\redcircle{1pt}}|\redcircle{1pt}}+0(1-p_{{\redcircle{1pt}}|\redcircle{1pt}})$
\end{tabular} &
$b<\frac{1+2gL}{1+2L}$\\
\hline
\end{tabular}
\end{center}

These two rationality constraints combine with one of the positivity constraints ($b>g$) and the social dilemma condition 3 ($L<B-1$) to define the valid region in which the signal-following norm is rational (see \cref{fig:phasediagram}b).
The most important remark about this phase diagram can be made using the more simple case of $g=0$, \textit{i.e.} the case where the signal statistics allow for completely avoiding the punishment from mutual exploitation. We also fix $B=3$, consistent with a widely used instance of the game of chicken (for which $L=0.5$; see black dot in \cref{fig:phasediagram}c). The system is now parametrized by the pair $(b,L)$. 
In this plane (\cref{fig:phasediagram}c), we expect that the maximum signal potential that can be realized while still ensuring it is rational to follow the signal-following norm should decrease as challenge level increases, and this is indeed the case. For a given challenge level, the highest average reward is obtained at this maximum rational signal potential (where the reward for $stop$ and $go$ are equal when the red signal \redcircle{2pt} is observed). 
We have represented reward relative to that of the best (mixed) Nash strategy and see that the correlated equilibrium performs better (values greater than 1) over a large area of the parameter space (up to around 1.8 times better for $L$ at its maximum of $B-1=2$).

Note that since $|\mathcal{A}|=|\mathcal{O}|=2$, there are 16 norms (4 possible prescriptions by 4 possible descriptions).
Of the 16 norms in the game of chicken and in this case that $g=0$ and $B=3$, the 4 norms with prescriptions that always \textit{stop} are null, leaving 12 rational norms (3 unique prescriptions by 4 unique descriptions). 
Alongside the default strategy, $\bm{P}_0$, of the mixed-strategy Nash equilibrium ($p_{stop}=1/(1+L)$ with average reward $\rho_{Nash}=(B+L)/(L+1)$), are the 3 unique prescriptions,
\begin{equation}
\bm{P}_1=    \begin{bmatrix}
        0&1\\
       1&0
    \end{bmatrix} \;,\;
    \bm{P}_2=\begin{bmatrix}
        1&0\\
       0&1
    \end{bmatrix} \;,\;\textrm{and}\;
    \bm{P}_3=    \begin{bmatrix}
        0&0\\
       1&1
    \end{bmatrix}.
\end{equation}
The identity policy, $\bm{P}_2$, is the prescription and description of the signal-following norm: just do what the signal says.  $\bm{P}_1$ is the anti-signal norm: do the opposite of what the signal says. Finally,  $\bm{P}_3$ prescribes that players always \textit{go}. For prescription indexing $n=0,1,2,3$, this gives for arbitrary pair event distribution $p$ (represented by $\bm{P}_{oo'}$) the following prescription payoff matrix $\bm{\Gamma}$ of average reward of strategy $n$ playing strategy $n'$ ($\Gamma_{nn'}=\rho(\pi,S)$ (\cref{eq:matrho}) each expressed as a function of $b$ and $L$,
\begin{equation}
    \bm{\Gamma}(b,L)=
    \begin{bmatrix}\frac{L + 3}{L + 1} & \frac{b - \left(b - 1\right) \left(L \left(L + 3\right) + 3\right) + 1}{2 \left(L + 1\right)} & \frac{- b + \left(b + 1\right) \left(L \left(L + 3\right) + 3\right) + 1}{2 \left(L + 1\right)} & \frac{1}{L + 1}\\\frac{L + 3}{L + 1} & - \frac{b}{2} - \frac{\left(L + 3\right) \left(b - 1\right)}{2} + \frac{1}{2} & b \left(L + \frac{3}{2}\right) + \frac{3}{2} & \frac{1}{2} - \frac{b}{2}\\\frac{L + 3}{L + 1} & \frac{3}{2} - \frac{b}{2} & \frac{5 b}{2} - \frac{\left(L + 3\right) \left(b - 1\right)}{2} + \frac{1}{2} & \frac{b}{2} + \frac{1}{2}\\\frac{L + 3}{L + 1} & - \frac{\left(L + 3\right) \left(b - 1\right)}{2} & \frac{\left(L + 3\right) \left(b + 1\right)}{2} & 0\end{bmatrix}
    \label{eq:gammamatrix}
\end{equation}



The highest scoring self-play norm (largest diagonal entry) is the signal following norm with reward $b(L-2)+3$ (the heat map in \cref{fig:phasediagram}c). This has maximums for fixed $L$  at the boundary of the rational region where the reward for $stop$ and $go$ are equal when \redcircle{2pt} is received. 
These rewards are significantly higher than the best Nash strategy.
Note that under this signal-following norm, the joint action distribution $p(a,a')=\bm{P}_2\bm{P}_{oo'}\bm{P}_2=\bm{P}_{oo'}$. A self-consistent interpretation of the correlating signal distribution is as the joint action distribution of another identical pair of agents or even some mean field (\textit{i.e.} population-averaged) pair-wise behaviour. This obviates an external environment to provide the signal. 

\subsection*{Replicator Dynamics of Social Norms}
Given $\bm{\Gamma}$ as a fitness matrix, replicator dynamics is a candidate fitness-based selection dynamics among types. Applied to norms, it provides a means by which a norm can establish itself in a many-agent system
\footnote{We focus on rational imitation by which agents meet in pairs, compare the payoffs for each's norm, and adopt the norm with the higher payoff. A slightly more general setting to imitation that is more in the spirit of the epistemic perspective outlined in the introduction is socalled \textit{pairwise comparison} in which the norm is adopted with some posterior probability given a noisy observation of the other player's reward. 
A posterior probability for unimodal noise model can be well-approximated by the logistic function,
\begin{align}
p(\Delta f)=\frac{1}{1+e^{-\beta\Delta f}}
\end{align}
where $\Delta f=|\Gamma_{nn'}-\Gamma_{n'n}|>0$ is the fitness difference of the two player's norms and $\beta$ is the strength of selection. This posterior
results exactly from a linear log-odds model with equal prior probabilities and coefficient parameter $\beta$ set as the strength of selection (e.g. noise strength in the observation model). This results in a replicator dynamics $\dot{x}_n/x_n=\tanh(\beta\Delta f_n)$ which reduces to \cref{eq:rep} in the weak selection limit (\textit{i.e.} $\beta\Delta f_n$ typically much less than 1) for which $\beta$ sets the speed of the dynamics.}.
It is obtained as the mean-field limit of the coupled birth-death process of $N$ populations of player types (defined by the prescription of the norm they abide by). The system is described in this limit  a continuous frequency vector $\bm{x}=(x_1,\dots,x_N)$, where $x_n\in[0,1]$ for $n\in\{1,\dots,N\}$ is the frequency of the $n$th population in the system and the components satisfy the normalization constraint $\sum_n x_n=1$. The limiting dynamics is exactly
\begin{align}
    \frac{\dot{x}_n}{x_n}=\Delta f_n \label{eq:rep}
\end{align}
with fitness deviations $\Delta f_n=f_n(\bm{x})-\bar{f}(\bm{x})$ from the mean fitness $\bar{f}(\bm{x})=\sum_n x_nf_n(\bm{x})$.\footnote{This dynamics is known as the generalized Lotka-Volterra equation in ecology, and is a continuous time version of the Price equation studied in the theory of biological evolution.} The fitness of a type when playing against the other type (including itself) is given by the expected payoff from playing in the system $f_n(\bm{x})=(\bm{\Gamma}\bm{x})_n$, where $\bm{\Gamma}$ is the fitness matrix. 
Direct evaluation shows that this dynamics respects the normalization constraint.
\Cref{eq:rep} can be written in matrix form
\begin{align}
    \dot{\bm{x}}=\mathrm{diag}(\bm{x})\left(\bm{\Gamma}\bm{x}-(\bm{x}^\top\bm{\Gamma}\bm{x})\bm{1}\right),\label{eq:matrixrep}
\end{align}
where the function $\mathrm{diag}(\bm{x})$ evaluates to a matrix with elements of the vector $\bm{x}$ on the diagonal and 0 elsewhere and $\bm{1}$ denotes a vector of ones of the same size as $\bm{x}$. We have written mean fitness explicitly, $\bar{f}(\bm{x})=\bm{x}^\top\bm{\Gamma}\bm{x}$.

\begin{figure}[tpb!]
\vskip 0.2in
\begin{center}
\blackcircle{2pt}$\xrightarrow{\makebox[4cm]{}}$\blackcircle{2pt}$\xrightarrow{\makebox[4cm]{}}$\blackcircle{2pt}

\includegraphics[width=0.3\textwidth]{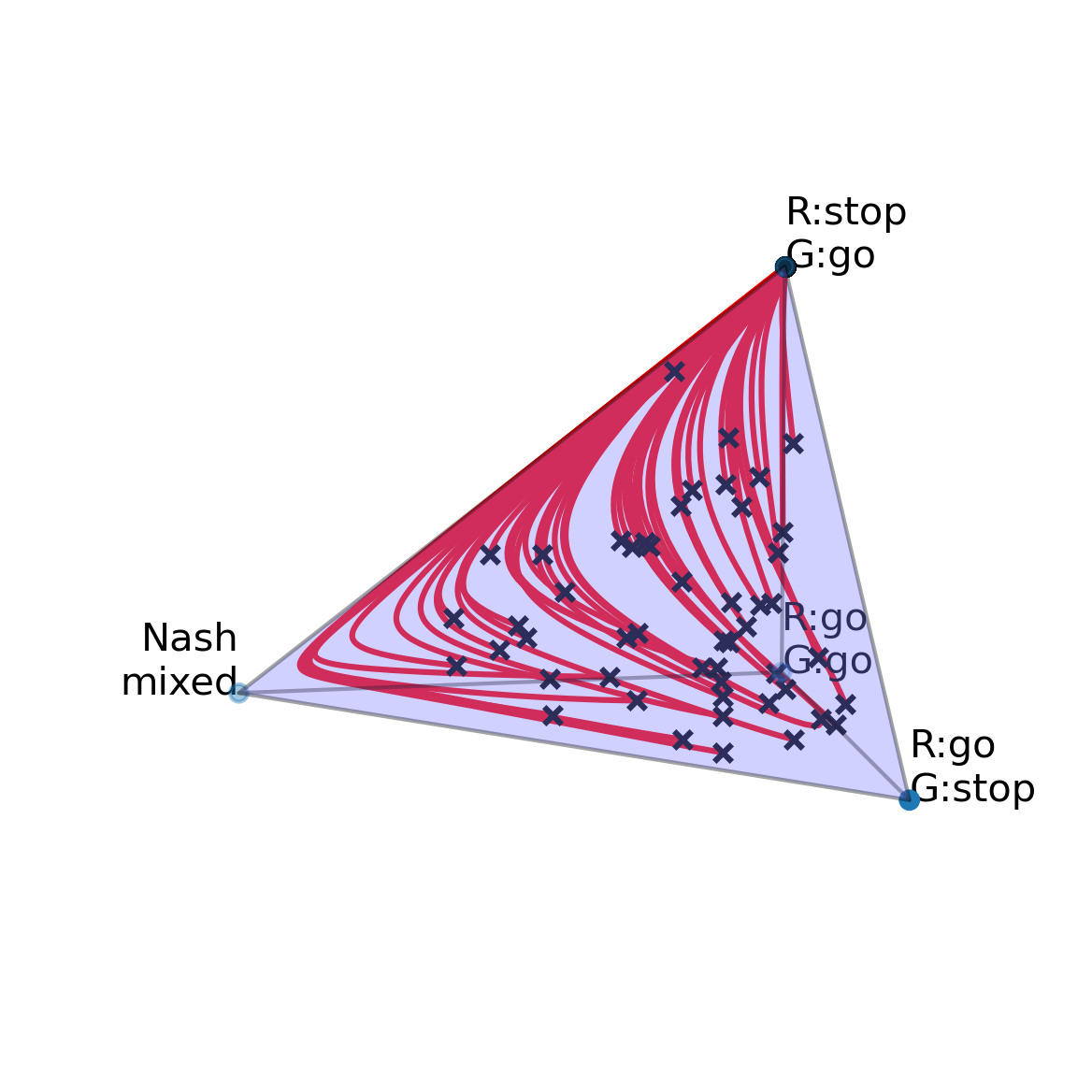}
\includegraphics[width=0.3\textwidth]{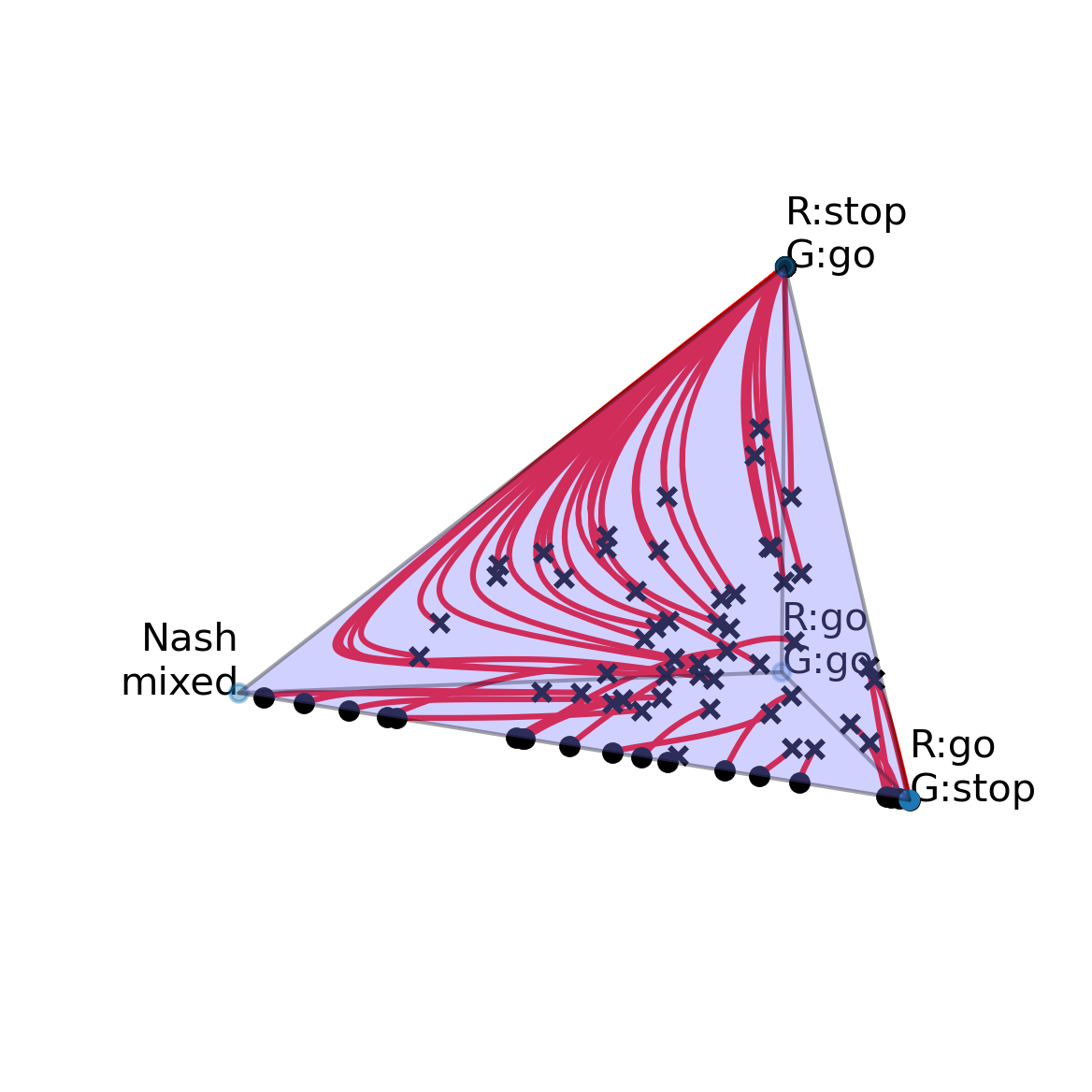}
\includegraphics[width=0.3\textwidth]{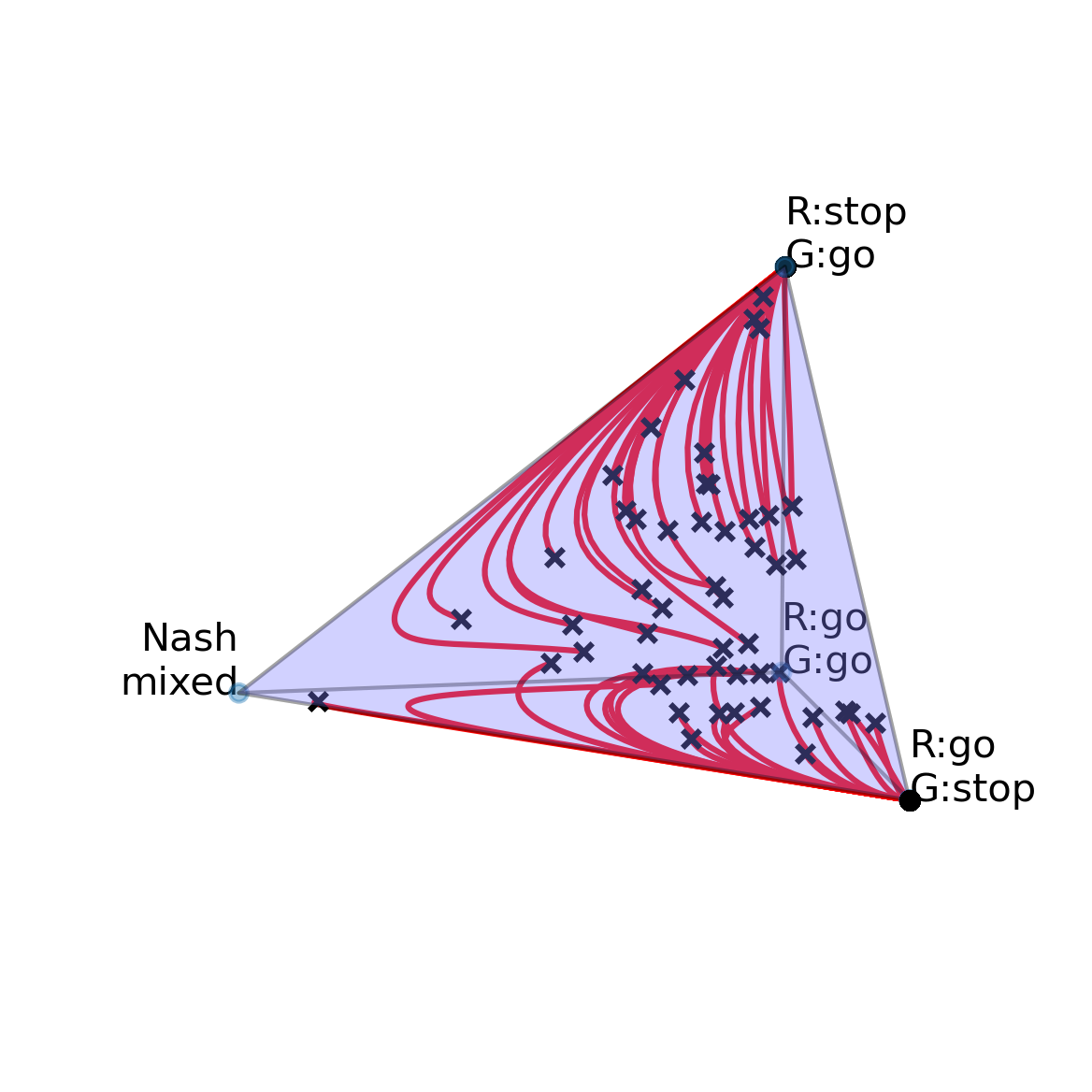}
\includegraphics[width=\textwidth]{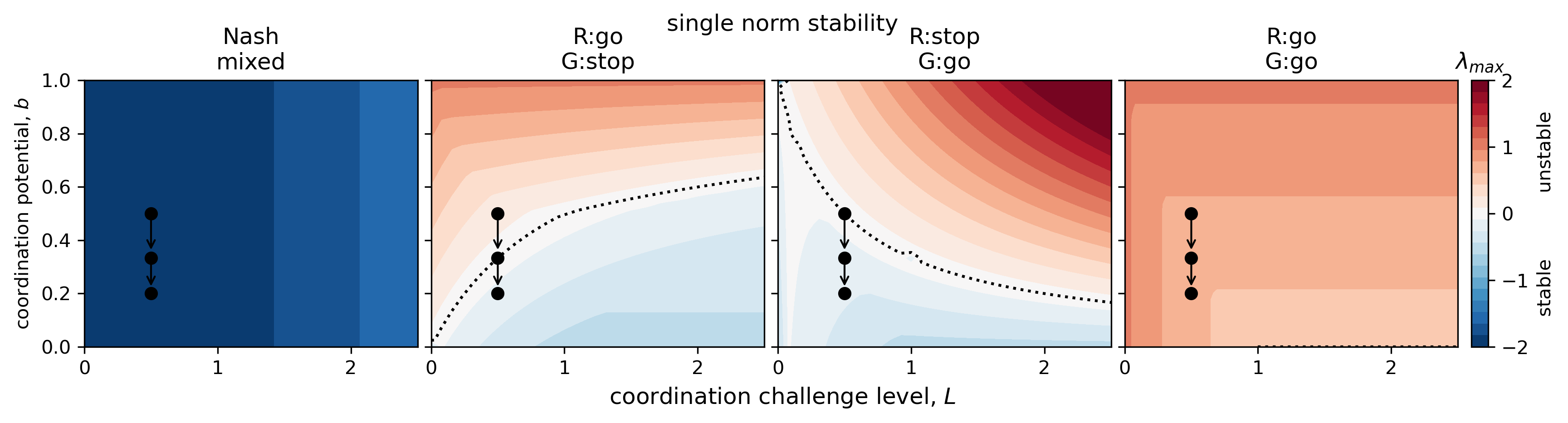}
\caption{\textbf{Replicator dynamics (\cref{eq:matrixrep}) for norms in the game of chicken.} \textbf{Top}: State space of norm frequencies in which 30 trajectories (red) are shown starting at frequencies sampled uniformily in the volume (cross markers) (left: $b=1/2$ , center: $b=1/3$, right: $b=1/5$). The trajectories end on the signal-following norm (black circle). Rational norms are denoted by the defining pair of $o:a$. 
\textbf{Bottom}: Eigenvalue spectra of the Jacobians at four single-norm fixed points over $(b,L)$. Dotted lines denote stability transitions ($\lambda_\mathrm{max}=0$).}
\label{fig:extraj}
\end{center}
\vskip -0.2in
\end{figure}

For the game of chicken example with $4\times4$ norm fitness matrix $\bm{\Gamma}$ given above \cref{eq:gammamatrix}, this is a dynamics on the 3-simplex, \textit{i.e.} the tetrahedron. Three qualitatively distinct example parametrizations of \cref{eq:matrixrep} for the optimal signal-following norm are given in \cref{fig:extraj}(top).

\paragraph{Solution Analysis}
The replicator dynamics represented by \cref{eq:matrixrep} has solutions that while complicated are tractable via standard dynamical systems analysis.
Fixed points, $\bm{x}^*$ are solutions to the matrix system
\begin{equation}
    \bm{\Gamma}\bm{x}=(\bm{x}^\top\bm{\Gamma}\bm{x})\bm{1}, 
\end{equation}
where $x_n$ is non-zero.
All states at which only a single norm $n$ is present (denoted $\hat{\bm{n}}$, \textit{i.e.}  $\hat{n}_i=\delta_{in}$) are fixed points\footnote{One can also show there are no interior equilibria}.
Given the flow map $g(\bm{x})$ (the righthand side \cref{eq:matrixrep}), stable fixed points are those fixed points $\bm{x}^*$ for which the maximum eigenvalue of the Jacobian of $g(\bm{x})$,  i.e. the $N \times N$ matrix of partial derivatives denoted by $\nabla_{\bm{x}} g(\bm{x})$, evaluated at $\bm{x}^*$ is non-positive. We derive a convenient matrix form here starting from the product rule,
\begin{align}
\left(\nabla_{\bm{x}} g(\bm{x})\right)_{ij}&=\frac{\partial g_i(\bm{x})}{\partial x_j}\\
&=\delta_{ij}((\bm{\Gamma}\bm{x})_j-\bm{x}^\top\bm{\Gamma}\bm{x})+x_i\left(\Gamma_{ij}-\left(2\Gamma_{jj}x_j+\sum_{i'\neq j}(\Gamma_{ji'}+\Gamma_{i'j}) x_{i'}\right)\right)\\
&=\delta_{ij}((\bm{\Gamma}\bm{x})_j-\bm{x}^\top\bm{\Gamma}\bm{x})+x_i\left(\Gamma_{ij}-((\bm{\Gamma}^\top+\bm{\Gamma})\bm{x})_j\right)\\
&=\delta_{ij}((\bm{\Gamma}\bm{x})_j-\bm{x}^\top\bm{\Gamma}\bm{x})+x_i\Gamma_{ij}-x_i(\bm{\Gamma}^\top\bm{x})_j-x_i(\bm{\Gamma}\bm{x})_j\\ 
\nabla_{\bm{x}} g(\bm{x})&=\mathrm{diag}(\bm{\Gamma}\bm{x}-(\bm{x}^\top\bm{\Gamma}\bm{x})\bm{1}) +\mathrm{diag}(\bm{x})\bm{\Gamma}-\bm{x}(\bm{\Gamma}^\top\bm{x})^\top-\bm{x}(\bm{\Gamma}\bm{x})^\top\\
&=\mathrm{diag}(\bm{\Gamma}\bm{x}-(\bm{x}^\top\bm{\Gamma}\bm{x})\bm{1}) +(\mathrm{diag}(\bm{x})-\bm{x}\bm{x}^\top)\bm{\Gamma}-\bm{x}\bm{x}^\top\bm{\Gamma}^\top\label{eq:jac}
\end{align}
where $\delta_{ij}=1$ if $i=j$ and 0 otherwise. 
%
Evaluating \cref{eq:jac} for single norm states $\hat{\bm{n}}$, the second term vanishes. Evaluating the first and third term, we obtain the Jacobian
\begin{align}
    (\nabla_{\bm{x}} g(\hat{\bm{n}}))_{ij}
    &=(\Gamma_{in}-\Gamma_{nn})\delta_{ij}-\Gamma_{jn}\delta_{in}\\
     \nabla_{\bm{x}} g(\hat{\bm{n}})&=\mathrm{diag}(\Gamma_{\cdot n}-\Gamma_{nn}\bm{1})-\hat{\bm{n}}(\Gamma_{\cdot n})^\top\label{eq:Jac}
\end{align}
This sparse matrix is composed of diagonal component $\Gamma_{in}-\Gamma_{nn}$ and an $n$th row component $-\Gamma_{jn}$. 
Given $\bm{\Gamma}$, the set of eigenvalues of the single state Jacobian matrix \cref{eq:Jac} can be computed numerically. 

For the game of chicken example ($\bm{\Gamma}$ given by \cref{eq:gammamatrix}), the result in the $(b,L)$ plane for the 4 single strategy states is shown in \cref{fig:extraj}(bottom). The always \textit{go} norm is always unstable. The mixed Nash strategy is always stable (and relatively strongly so), but its attractor basin extends over a negligibly small fraction of the state space (frequency perturbations of magnitude $10^{-3}$ were sufficient to knock the system out of that equilibrium). Across a path in $(b,L)$ on which the coordination potential is lowered, the system transitions from being the attractor basin of the signal-following norm taking up the vast majority of the state space (\cref{fig:extraj}top,left) to a more balanced multistability in which the attractor basin of the anti-signal following norm also takes up a significant fraction of the state space volume (\cref{fig:extraj}top,right). Note that at the critical value of $b$ at which the anti-signal following norm becomes stable, a significant state space volume attracts to a population mixture of that norm and the mixed Nash strategy (\cref{fig:extraj}top,middle). 

\section*{Discussion}
We presented a detailed account of how social norms can emerge via the use of correlated equilibria.

\begin{ack}
    MPT acknowledges helpful conversations with Bryce Morksy and Matthew Riemer.
\end{ack}

\bibliographystyle{unsrtnat}
\bibliography{references}

@article{Morsky2019,
author = {Bryce Morsky  and Erol Akçay },
title = {Evolution of social norms and correlated equilibria},
journal = {Proceedings of the National Academy of Sciences},
volume = {116},
number = {18},
pages = {8834-8839},
year = {2019},
doi = {10.1073/pnas.1817095116},
URL = {https://www.pnas.org/doi/abs/10.1073/pnas.1817095116},
eprint = {https://www.pnas.org/doi/pdf/10.1073/pnas.1817095116}}

@Article{Hilbe2011,
author={Hilbe, Christian},
title={Local Replicator Dynamics: A Simple Link Between Deterministic and Stochastic Models of Evolutionary Game Theory},
journal={Bulletin of Mathematical Biology},
year={2011},
month={Sep},
day={01},
volume={73},
number={9},
pages={2068-2087},
issn={1522-9602},
doi={10.1007/s11538-010-9608-2},
url={https://doi.org/10.1007/s11538-010-9608-2}
}

@article{Foster1997,
title = {Calibrated Learning and Correlated Equilibrium},
journal = {Games and Economic Behavior},
volume = {21},
number = {1},
pages = {40-55},
year = {1997},
issn = {0899-8256},
doi = {https://doi.org/10.1006/game.1997.0595},
url = {https://www.sciencedirect.com/science/article/pii/S0899825697905959},
author = {Dean P. Foster and Rakesh V. Vohra}
}

@article{Hart2000,
author = {Hart, Sergiu and Mas-Colell, Andreu},
title = {A Simple Adaptive Procedure Leading to Correlated Equilibrium},
journal = {Econometrica},
volume = {68},
number = {5},
pages = {1127-1150},
doi = {https://doi.org/10.1111/1468-0262.00153},
url = {https://onlinelibrary.wiley.com/doi/abs/10.1111/1468-0262.00153},
eprint = {https://onlinelibrary.wiley.com/doi/pdf/10.1111/1468-0262.00153},
year = {2000}
}

@article{Aumann1987,
 ISSN = {00129682, 14680262},
 URL = {http://www.jstor.org/stable/1911154},
 author = {Robert J. Aumann},
 journal = {Econometrica},
 number = {1},
 pages = {1--18},
 publisher = {[Wiley, Econometric Society]},
 title = {Correlated Equilibrium as an Expression of Bayesian Rationality},
 urldate = {2023-05-08},
 volume = {55},
 year = {1987}
}

@article{
Macy2002,
author = {Michael W. Macy  and Andreas Flache },
title = {Learning dynamics in social dilemmas},
journal = {Proceedings of the National Academy of Sciences},
volume = {99},
number = {suppl\_3},
pages = {7229-7236},
year = {2002},
doi = {10.1073/pnas.092080099},
URL = {https://www.pnas.org/doi/abs/10.1073/pnas.092080099},
eprint = {https://www.pnas.org/doi/pdf/10.1073/pnas.092080099}}

@inproceedings{Leibo2017,
author = {Leibo, Joel Z. and Zambaldi, Vinicius and Lanctot, Marc and Marecki, Janusz and Graepel, Thore},
title = {Multi-Agent Reinforcement Learning in Sequential Social Dilemmas},
year = {2017},
publisher = {International Foundation for Autonomous Agents and Multiagent Systems},
address = {Richland, SC},
booktitle = {Proceedings of the 16th Conference on Autonomous Agents and MultiAgent Systems},
pages = {464–473},
numpages = {10},
location = {S\~{a}o Paulo, Brazil},
series = {AAMAS '17}
}

@article{Axelrod,
 ISSN = {00030554, 15375943},
 URL = {http://www.jstor.org/stable/1960858},
 author = {Robert Axelrod},
 journal = {The American Political Science Review},
 number = {4},
 pages = {1095--1111},
 publisher = {[American Political Science Association, Cambridge University Press]},
 title = {An Evolutionary Approach to Norms},
 urldate = {2023-06-08},
 volume = {80},
 year = {1986}
}

@inproceedings{Anagnostides,
author = {Anagnostides, Ioannis and Farina, Gabriele and Kroer, Christian and Celli, Andrea and Sandholm, Tuomas},
title = {Faster No-Regret Learning Dynamics for Extensive-Form Correlated and Coarse Correlated Equilibria},
year = {2022},
isbn = {9781450391504},
publisher = {Association for Computing Machinery},
address = {New York, NY, USA},
url = {https://doi.org/10.1145/3490486.3538288},
doi = {10.1145/3490486.3538288},
booktitle = {Proceedings of the 23rd ACM Conference on Economics and Computation},
pages = {915–916},
numpages = {2},
location = {Boulder, CO, USA},
series = {EC '22}
}

@inproceedings{Greenwald,
author = {Greenwald, Amy and Hall, Keith},
title = {Correlated-Q Learning},
year = {2003},
isbn = {1577351894},
publisher = {AAAI Press},
booktitle = {Proceedings of the Twentieth International Conference on International Conference on Machine Learning},
pages = {242–249},
numpages = {8},
location = {Washington, DC, USA},
series = {ICML'03}
}

@article{Gintis,
author = {Herbert Gintis},
title ={Social norms as choreography},
journal = {Politics, Philosophy \& Economics},
volume = {9},
number = {3},
pages = {251-264},
year = {2010},
doi = {10.1177/1470594X09345474},

URL = { 
        https://doi.org/10.1177/1470594X09345474
    
},
eprint = { 
        https://doi.org/10.1177/1470594X09345474
    
}
}

@article{
Tornberg,
author = {Petter Törnberg },
title = {How digital media drive affective polarization through partisan sorting},
journal = {Proceedings of the National Academy of Sciences},
volume = {119},
number = {42},
pages = {e2207159119},
year = {2022},
doi = {10.1073/pnas.2207159119},
URL = {https://www.pnas.org/doi/abs/10.1073/pnas.2207159119},
eprint = {https://www.pnas.org/doi/pdf/10.1073/pnas.2207159119}}

@article{Vinitsky,
author = {Vinitsky, Eugene and K\"{o}ster, Raphael and Agapiou, John P and Du\'{e}\~{n}ez-Guzm\'{a}n, Edgar A and Vezhnevets, Alexander S and Leibo, Joel Z},
title = {A Learning Agent That Acquires Social Norms from Public Sanctions in Decentralized Multi-Agent Settings},
year = {2023},
issue_date = {April-June 2023},
publisher = {Sage Publications, Inc.},
address = {USA},
volume = {2},
number = {2},
url = {https://doi.org/10.1177/26339137231162025},
doi = {10.1177/26339137231162025},
journal = {Collective Intelligence},
month = {apr},
numpages = {14}
}

@article{Koster,
author = {Raphael Köster  and Dylan Hadfield-Menell  and Richard Everett  and Laura Weidinger  and Gillian K. Hadfield  and Joel Z. Leibo },
title = {Spurious normativity enhances learning of compliance and enforcement behavior in artificial agents},
journal = {Proceedings of the National Academy of Sciences},
volume = {119},
number = {3},
pages = {e2106028118},
year = {2022},
doi = {10.1073/pnas.2106028118},
URL = {https://www.pnas.org/doi/abs/10.1073/pnas.2106028118},
eprint = {https://www.pnas.org/doi/pdf/10.1073/pnas.2106028118}}

\newpage
\appendix
\section*{Appendix}
\subsection*{Observing the agent population to close the loop}
We detailed an evolutionary dynamics of social norms for fixed, given statistics of the signal over a defined signal space that is used to correlate behaviours in each norm. 
Here, motivated by the large amounts of social information in modern human experience, we instead consider the natural case that signal statistics are in fact evolving according to some dynamics derived from partial observations of the population-level behaviour itself. This self-referential choice obviates the need for an external environment to provide correlated signals by leveraging the correlations inherent in the joint action of the many-agent system. 

Consider a partially observable $N$-agent Markov game ($N$ even) in which agents are paired according to a (possibly stochastic) pairing $\mathcal{P}$ at each time step to play a 2-player game and where the current state is the joint action $\bm{a}=(a_1,\dots,a_N)$ from the previous time step. One simple type of partial observation is the pair of actions from a game from the last round.\footnote{An alternative choice is that agents observe noisy versions of the population-averaged policy (the pair actions over all games played in the previous time step).} Whether or not pairs of agents abide by their subscribed norm however depends on whether it is rational to do so, otherwise they play (in our setting) the optimal Nash equilibrium. We assume a similar evolutionary dynamics that drives the frequencies of norm and signal pairs used in a population towards those that provide high average payoff, thus allowing the emergence of a correlated equilibrium in the same way as in the previous section.
The payoff matrix $\bm{\Gamma}$ now depends on these changing observation statistics
\begin{align}
    \Gamma_{nn'}(t)&=\mathrm{tr}(\bm{R}\bm{P}^{(nn')}_{aa'}(t))  
\end{align}
where 
\begin{align}
\bm{P}^{(nn')}_{aa'}(t)=\bm{P}^{(n')\delta}\bm{P}^{(nn')}_{oo'}(t)(\bm{P}^{(n)\delta})^\top
\end{align}
is the joint action distributions for pairings of agents with norm $n$ and $n'$, where $\bm{P}^{(n)\delta}$ is the conditional best-response strategy 
\begin{equation}
    \bm{P}^{(n)\delta}=\delta_{nn'}\bm{P}^{(n)}+(1-\delta_{nn'})\mathrm{diag}(\bm{p}_{\mathrm{Nash}})
\end{equation}
where $\delta_{nn'}=1$ if the norm $(\bm{P}^{(n)},\bm{D}^{(n)})$ is rational given $n'$ and $\bm{P}^{(nn')}_{oo'}(t)$ and 0 otherwise. $\bm{P}^{(n')\delta}$ is defined similarly. Finally, observations are simply empirical frequencies of actions played in the previous round of games between the two types of players, $\bm{P}^{(nn')}_{oo'}(t)=\hat{\bm{P}}^{(nn')}_{aa'}(t-\mathrm{d}t)$, where the latter is a finite-size estimator, \textit{e.g.} for $|\mathcal{A}|=2$, $\hat{\bm{P}}^{(nn')}_{aa'}(t)=\frac{1}{N_{nn'}}\sum_{(i,j)\in\mathcal{P}_{nn'}}\begin{bsmallmatrix}(1-a_i)(1-a_j) &(1-a_i)a_j\\a_i(1-a_j) & a_ia_j\end{bsmallmatrix}$, where we use the action representation $0\sim stop$ and $1\sim go$ and $\mathcal{P}_{nn'}\subset\mathcal{P}$ is the subset of the pairings made for games in time step, $t$, from players of the two types, and $N_{nn'}=|\mathcal{P}_{nn'}|$ is the number of these pairings. When $N_{nn'}$ is large (so that finite-size fluctuations are negligible), $\hat{\bm{P}}^{(nn')}_{aa'}(t)\to \bm{P}^{(nn')}_{aa'}(t)$. By construction then, the single norm state of the signal-following norm is stable with respect to this closed-loop multi-agent dynamics. 

\subsection*{Adding the assumption of partisan interactions leads to partisan sorting}
Polarization in existing models on the effects of social media on opinion dynamics is driven by an absorbing dynamics that homogenizes opinions within groups. These models achieve differing opinions across groups with high probability by the low probability under uniform sampling of many opinions that different groups land on the same opinion. This mechanism is unrealistic in a few ways: the group opinions are uncorrelated with unrealistic parameter dependence, e.g. the fewer opinions the less polarization. A more compelling hypothesis not accounted for in these models is that the in-group homogenized opinions are \textit{anti-correlated} across groups. Here, we propose to induce this anti-correlation as a consequence of a correlated equilibria that choreographs conflict.

Consider the following setting that implements the idea that assumed partisanship leads to partisan sorting. Each player assumes there are two agent types and know which type with which it identifies more.
Players then perceive reward functions that differ for like and unlike pair interactions with other agents, $\bm{R}=\begin{bsmallmatrix}
    B &2\xi-1\\B+L&0
\end{bsmallmatrix}$, where $\xi=1$ for opponents of the same type (with whom the player then plays a game of chicken) and $\xi=0$ for opponents of the other type (with whom the player then plays a game of stag hunt so that being exploited is now more costly than mutually exploiting one another). The optimal player strategy thus depends on the opponent's type, $\xi$, which the player strives to infer in rational pursuit of higher reward. The basis of this inference (the computation leading to the estimate $\hat{\xi}$) lies in the inferred similarity of opinions with the opponent from partial observations of their opinions. We implement this by augmenting each player with a private belief space (represented relative to the population-averaged beliefs) as a vector $\bm{d}\in\mathbb{R}^D$ ($D$ could be interpreted as the number of topics). Opinion similarity is defined by an opinion similarity function, $S(\bm{d},\bm{d}')$. Here, we assume that in pursuit of inferring similarity players binarize their opinions relative to the average and use the normalized overlap $S=S(\bm{d},\bm{d}')=\tfrac{1}{D}\Theta(\bm{d})^\top\Theta(\bm{d}')\in[0,1]$ to define similarity where $\Theta$ is applied element-wise ($\Theta(x)=1$ if $x>0$, 0 otherwise) and set $\xi=\xi(S)=\Theta(S-1/2)$. An opponent's opinion $\bm{d}'$ is partially observable, with single components revealed in single games. Prior assumptions about the distributions of opinions can strongly influence the inference of $\xi$. Under the prior assumption of non-overlapping variability of two types, $\hat{\xi}=0$ upon the observation of any deviating opinion component. Alternatively, the maximally uninformative prior leads to a posterior with significant residual uncertainty on $S$ from single observations that provide only a component of the opinion vector.


The different reward matrix for each pairing type drives the selection of a pair of action pair distributions via the joint-action derived signal statistics, $\bm{P}^+_{oo'}$ and $\bm{P}^-_{oo'}$. We propose that the same dynamics that leads to a correlated equilibria for the policies conditioned on playing opponents from the same population, will lead to a suboptimal equilibria when conditioned on playing against those of the different population that is characterized by choreographed conflict rather than cooperation. We also aim to show that the dependence of $\bm{R}$ on the population label can in fact be based solely on the player's estimate and thus no ground truth, type-specific populations need exist. The latter emphasizes that it is the interaction dynamics out of which the conflicted populations emerge. This is in line with recent proposals for partisan sorting via conflict alignment~\citep{Tornberg}. Under which conditions the set of local assignments are mutually consistent is then a focal question when aiming to understand the stationary behaviour of the system.

\end{document}